\documentclass[twocolumn,showpacs,preprintnumbers,amsmath,amssymb,floatfix]{revtex4}

\usepackage{graphicx}
\usepackage{dcolumn}
\usepackage{bm}
\usepackage{hyperref}
\usepackage{color}
\usepackage{epsfig}

\newcommand{\nid}{\noindent}

\newcommand{\beq}{\begin{equation}}
\newcommand{\eeq}{\end{equation}}
\newcommand{\ds}{\displaystyle}

\begin{document}

\preprint{Preprint}

\title[A source-free integration method]{A source-free integration method for black hole perturbations\\ and self-force computation: radial fall\footnote{This arXiv version differs from the one published in Phys. Rev. D for the use of British English and other minor editorial differences.}}
\author{Sofiane Aoudia}

\affiliation{Max Planck Institut f\"ur Gravitationphysik, A. Einstein, Golm\\
1 Am M\"uhlenberg 14476 Golm, Deutschland\\
E-mail: aoudia@aei.mpg.de}

\author{Alessandro D.A.M. Spallicci}

\affiliation{
Universit\'e d'Orl\'eans, 
Observatoire des Sciences de l'Univers en r\'egion Centre OSUC\\
LPC2E-CNRS $\!\!\!\!$ 3A $\!\!\!\!$ Av. $\!\!\!\!\!\!\!$ Recherche $\!\!\!\!$ Scientifique $\!\!\!\!$ 45071 $\!\!\!\!\!$ 
Orl\'eans, $\!\!\!\!\!$ France\\
E-mail: spallicci@cnrs-orleans.fr}

\date{7 March 2011}

\begin{abstract}
Perturbations of Schwarzschild-Droste black holes in the Regge-Wheeler gauge benefit from the availability of a wave equation and from the gauge invariance of the wave function, but lack smoothness.   
Nevertheless, the even perturbations belong to the C\textsuperscript{0} continuity class, if the wave function and its derivatives satisfy specific conditions on the discontinuities, known as jump conditions, at the particle position.   
These conditions suggest a new way for dealing with finite element integration in the time domain.  
The forward time value in the upper node of the $(t, r^*$) grid cell is obtained by the linear combination of the three preceding node values and of analytic expressions based on the jump conditions. The numerical integration does not deal directly with the source term, the associated singularities and the potential. This amounts to an indirect integration of the wave equation.   
The known wave forms at infinity are recovered and the wave function at the particle position is shown. 
In this series of papers, the radial trajectory is dealt with first, being this method of integration applicable to generic orbits of EMRI (Extreme Mass Ratio Inspiral). 
\end{abstract}


\maketitle

\section{Introduction}

It is currently affirmed that the core of most galaxies
host supermassive black holes, on which stars and compact objects in the neighbourhood inspiral and plunge-in. The EMRI (Extreme Mass Ratio Inspiral) sources are characterised by a huge number of parameters that, when spanned over a large period, produce a {huge number of templates for the LISA (Laser Interferometer Space Antenna) project \cite{li}. Thus, matched filtering (the superposition of the expected and received signals)\cite{sasc03} is aided by stochastic methods \cite{po09}; in alternative, other more robust methods based on covariance or on time and frequency analysis are investigated \cite{sasc03, gajo07, gamawe08a, gamawe08b, co10, bagapo09}. Furthermore, if the signal from a capture is not individually detectable, it still may contribute to the statistical background \cite{bacu04,racu07}. Detection strategies are to be tested by the LISA simulators \cite{coru03, va05, peauhajeplpirevi08}.

Challenges by EMRIs arise increasing interest 
amongst astrophysicists and those studying data analysis, but also for theorists. Relativists, especially, have been drawn to study EMRIs for the determination of the motion of the captured mass and because of the impact that back-action has on the gravitational wave forms (see \cite{blspwh10} where a comprehensive introduction to these topics is given).

The captured compact object may be compared to a small mass $m$ perturbing the background spacetime curvature of a large mass 
$M$ and generating gravitational radiation. Perturbation methods from the seminal work of Regge and Wheeler, henceforth RW \cite{rewh57}, were an obvious aid for studying the case we are dealing with. Approaches based on numerical relativity (NR) appear not be best suited, nor do the post-Newtonian (pN) expansions: the former  
for the two-scale problem (one small scale for the neighbourhood of the particle, one large scale for the radiation at infinity); the latter for the small field and low velocity constraints.} To EMRIs, nevertheless, Damour, Nagar and co-workers
\cite{nadata07, dana07, bena10, da09b}, and Yunes {\it et al.} \cite{yubuhumipa10} have applied the EOB (effective one body)
method, though based on the pN approach. Furthermore, 
both pN and perturbation methods are found to be in agreement in their common domain of applicability by Blanchet {\it et al.} 
 \cite{bldeltwh10a, bldeltwh10b}, as well as EOB and perturbation methods have shown their synergy thanks to Barack {\it et al.} \cite{badasa10}. Finally, it is to be mentioned that NR analysis is slowly progressing towards unequal mass ratios \cite{sp10}. 
       
\subsection{Perturbation methods}

EMRIs have revived the interest for the perturbative relativistic two-body problem, 
 which was investigated with a semi-relativistic approach some 40 years ago by Ruffini and Wheeler \cite{ruwh71a,ruwh71b}, and 
then more fully by Zerilli \cite{ze69, ze70a, ze70b, ze70c}.  
These authors treated the source of the perturbations in the form of a radially falling particle. Zerilli's results opened the way for others to study to first order the orbital motion of the captured mass in a fully, although linearised, relativistic regime.
Work done up until the end of the '90s
involved the captured star moving on the geodesic of the background field and being unaffected by its own mass and the emitted radiation.

We shall refer to the first solution of the Einstein equation as the work of Droste \cite{dr15, dr16a, dr16b} and Schwarzschild \cite{sc16}, collectively as SD, this double authorship having been justified by the historians, as recalled by Rothman \cite{ro02}. 
Much research has been devoted to the study of SD black hole perturbations in the RW gauge, first in a vacuum \cite{rewh57}, and then by Zerilli in the presence of a source \cite{ze69, ze70a, ze70b, ze70c}. 
An outflow of publications by Davis, Press, Price, Ruffini and Tiomno \cite{daruprpr71, daru72, daruti72, ru73a, ru73b, ru78}, but also by individual 
scholars like Chung \cite{ch73}, Dymnikova \cite{dy80} appeared in the '70s. It was followed  by a slowdown in the '80s and '90s, with the exception of the forerunners of the Japanese school as Tashiro and Ezawa \cite{taez81}, Nakamura with Oohara and Koijma \cite{naooko87} or with Shibata \cite{shna92}.
The aim of all these authors was to analyse especially the amplitude and spectrum of the radiation emitted by the falling mass.
Simulations were performed in frequency domain with the particle falling from infinity. 
The numerical accuracy was recently improved by Mitsou \cite{mi10}. 
It was in 1997, that a fall from finite distance was analysed by Lousto and Price \cite{lopr97a}. The latter two authors were the first to use a finite difference scheme in time domain \cite{lopr97b}. 
Following on this work, Martel and Poisson \cite{mapo02} were able to parametrise the initial conditions reflecting past motion of the particle and resulting into an initial amount of gravitational wave energy. All of this work was carried out in the RW gauge. 

It is less than 15 years since we began to have methods for evaluating the back-action for point masses in strong fields thanks to two concurring situations.
On one hand, the theorists progressed in understanding radiation reaction and obtained formal prescriptions for its determination.
The solution was brought by the self-force equation, first by Mino, Sasaki and Tanaka \cite{misata97}, Quinn and Wald 
\cite{quwa97}, and later by Detweiler and Whiting \cite{dewh03}, all various approaches yielding the same formal expression. 
On the other hand, for the above mentioned detection of captures, the requirements from LISA posed constraints on the tolerable amount of phase-shift on the wave forms, caused by radiation reaction. This impulse - project oriented - 
compelled the researchers to turn their efforts in finding an efficient and clear implementation of the prescription made by the theorists. The development was again pursued in perturbation theory. This time, though, the small mass $m$ is correcting the geodesic equation of motion on a fixed background via a factor ${\cal O}(m)$. 

Actually, the advances in the field by e.g. Warburton and Barack \cite{waba10} and in radiation gauge 
by Keidl {\it et al.} \cite{KeShFrKiPr10}
have recently tackled the self-force in Kerr geometry \cite{ke63}. 
Such initial results in Kerr geometry do not mean that all major issues in non-rotating black holes have been solved. For instance,  
the self-consistent evolution of an orbit, along the lines suggested by Gralla and Wald \cite{grwa08}, is not yet applied even to 
the simpler case of non-rotating black holes. 

\subsection{A different integration approach}

The complexity in assessing the continuity of the perturbations (and the associated numerical computation of derivatives) at the position of the particle, has led Barack and Lousto \cite{balo05}, among other motivations such as  
the compatibility of the self-force to the (Lorenz-FitzGerald-de Donder) harmonic gauge \cite{lo67, hu91, dd21}, to convey their efforts to this gauge. This alternative enterprise is by no means priceless, the price being the unavailability of a wave equation like in the RW gauge and of the gauge invariance of the wave function, as determined by Moncrief \cite{mo74}. In this context, the recent investigations of a gauge independent form of the self-force \cite{gr10} may change sensibly the trade-off on the gauge choice.  

The choice of a gauge should be suited to the type of problem under scrutiny and thus we do not adopt a position on which gauge should be preferred {\it a priori}. We find of some interest to revert a weakness of the RW gauge, namely the lack of smoothness, into a building tool of a new integration method. 

Indeed, we propose herein a finite element method of integration based on the jump conditions that the wave function and its derivatives have to satisfy for the SD black hole perturbations to be continuous at the position of the particle. We first deal with the radial trajectory and the associated even parity perturbations, while in a forthcoming paper \cite{aosp11} we shall present the circular and eccentric orbital cases, referring thus to both odd and even parity perturbations. 

Herein, we present a first order method and show that it suffices to acquire a well-behaved wave function at infinity and at the particle position. However, a higher order scheme has been derived and tested \cite{rispaoco11}.

The main feature of the indirect (-I-) method consists in avoiding the explicit integration of the wave equation (the source term with the associated singularities and the potential) whenever the grid cells are crossed by the particle. Indeed, the information on the wave equation is implicitly given by the jump conditions. Conversely, for cells not crossed by the particle,  we retain the classic approach for integrating
wave equations \cite{lopr97b,mapo02}, henceforth named LPMP.

For the computation of the back-action, the -I- method ensures a well behaved wave function at the particle position, since the approach is governed by the theoretical values of the jump conditions. 

\subsection{Structure of the paper}

The structure of the paper is as follows. Section II is a brief reminder of the formalism on perturbations for a SD black hole in the RW gauge. In Sec. III, we give an overview about the jump conditions on the wave function and its derivatives. In Sec.  IV, we introduce the scheme by detailing one of the four physical cases that a particle crossing the mesh cell during its infall may encounter, while for the other three similar cases we simply provide the final result. In Sec. V some wave forms at infinity are shown, discussed and compared to those obtained with the LPMP method. Furthermore, the theoretical and numerical jump conditions at the position of the particle are displayed and commented. In the conclusions, Sec. VI, the perspectives are addressed. 

Geometric units ($G = c = 1$) are used, unless stated otherwise. The metric signature is $(-, +, +, +)$.

\section{Black hole perturbations}

The Zerilli-Moncrief (ZM) equation rules even-parity waves in the presence of a source:  a freely falling point particle,
that generates a perturbation for which the difference from the 
SD geometry is small. The energy-momentum 
tensor $T _{\mu\nu}$ is given 
by the integral of the world line of the particle, the integrand containing  
a four-dimensional invariant $\delta$ Dirac distribution for the representation of the point particle trajectory. 
The vanishing of the covariant divergence of $T _{\mu\nu}$ is guaranteed by 
the world line being a 
geodesic in the background SD geometry, represented by the metric tensor $g_{\mu\nu}$. Finally, the complete description of 
the emitted gravitational waves is given by the symmetric tensor $h_{\mu\nu} \ll g_{\mu\nu}$. 

The formalism can be summarised as follows. Any symmetric covariant tensor can be expanded in spherical harmonics \cite{ze70b}.
Because of the spherical symmetry of the SD field, the 
linearised field equations are cast in the form of a rotationally  
invariant operator on $h_{\mu\nu}$. This term is equated to the energy-momentum tensor, 
also expressed in spherical tensorial harmonics. That is: 

\beq
Q[h_{\mu\nu}] \propto T_{\mu\nu}[\delta (r_u)]~~,
\label{qt}
\eeq
where the $\delta(r_u)$ Dirac distribution represents the point particle unperturbed trajectory $r_u$.  
The rotational invariance is used to separate out the angular variables in the 
field equations. For the spherical symmetry of the 2-dimensional manifold on which $t,r$ are constants under rotation in the $\theta,\phi$ sphere, the ten components of the perturbing symmetric tensor transform like three scalars, two vectors and one tensor:
 
\[
h_{tt}, h_{tr}, h_{rr}
~~~~~~~~
~~~
(h_{t\theta}; h_{t\phi}), (h_{r\theta}; h_{r\phi})
~~~~~~~~
~~~
\left(\begin{matrix}
h_{\theta\theta} & h_{\theta\phi} \cr 
h_{\phi\theta} & h_{\phi\phi}
\end{matrix}\right )~~.
\]

In the Regge-Wheeler-Zerilli formalism, the source term for the odd perturbations vanishes for the radial trajectory and, given the rotational invariance through the azimuthal angle, only the index referring to the polar or  latitude angle survives. The even perturbations, going as $(-1)^l$, 
are expressed by the following matrix:

\begin{widetext}
{\small
\beq
h_{\mu\nu} = 
\left(
\begin{matrix}
\left (1-\frac{\displaystyle 2M}{\displaystyle r }\right )H^l_{0}Y^l  
&  H^l_{1}Y^l 
&  h^l_0 Y^l_{,\theta} 
& h^l_0 Y^l_{,\phi}\cr
sym 
&  \left (1-\frac{\displaystyle 2M}{\displaystyle r}\right )^{-1}H^l_{2}Y^l 
&  h^l_1 Y^l_{,\theta}   
& h^l_1 Y^l_{,\phi} 
\cr                        
sym
&  sym
&  
r^2 \left [ K^l Y^l + G^l Y^l_{,\theta\theta}
\right ] 
& r^2 G^l
\left (Y^l_{,\theta\phi} 
- \cot\theta Y^l_{,\phi} 
\right ) 
\cr
sym 
&  sym 
&  sym 
&  r^2\sin^2\theta\left[ 
K^l + G^l\left(\frac{\ds Y^l_{,\phi\phi}}{\ds\sin^2\theta}
+ \cot \theta Y^l_{,\theta}
\right ) \right ] 
\end{matrix}
\right )~~, 
\label{eq:rweven}
\eeq
}
\end{widetext}
where $H^l_0, H^l_1, H^l_2, h^l_0, h^l_1, K^l, G^l$ are functions of $(t, r)$ and the spherical harmonics $Y^l$ obviously of $(\theta,\phi)$.  
Two formalisms describe the evolution of the perturbations in terms of a
single wave function and a single wave equation. One, due to Moncrief \cite{mo74}, gauge invariant and developed about 
a 3-geometry of a t = constant hypersurface, refers solely to the $H^l_2, K^l, G^l, h^l_1,$ perturbations.  
The other, due to Zerilli  \cite{ze69, ze70a, ze70b, ze70c}, uses the RW gauge $G^l = h^l_0 = h^l_1 = 0$. In the RW gauge, the 
Moncrief wave function is given by: 
\beq
\Psi_l (t,r)= \frac {r}{\lambda +1}
\left[ K^l+\frac{r-2M}{\lambda r+3M}\left(
H_2^l-r\frac{\partial K^l}{\partial r}\right) \right],
\label{psidef}
\eeq
where the Zerilli \cite{ze70a} normalisation is used for $\Psi_l $. 
The dimension of the wave function is such that the energy is proportional to $\int_0^\infty  \dot\Psi^2\,dt $. 
Finally,
the wave equation is given by:  
\beq
\frac {\partial ^{2} \Psi^l(t,r)}{\partial r^{*2}} - \frac {\partial ^{2} \Psi^l(t,r)}{\partial t^2} -
V^{l}(r)\Psi^l (t,r) = S^{l}(t,r)~~,  
\label{eq:rwz*}
\eeq
\\
{\nid where $r^* = r + 2M\ln (r/2M - 1)$ is the tortoise coordinate. The potential $V^{l}(r)$ is expressed by:} 
\begin{widetext}
\beq
V^{l}(r) = \left (1 - \frac{2M}{r}\right )
\frac {2\lambda^{2}(\!\lambda\!+\!1\!) r^{3}\!\!+\! 6 \lambda^{2}Mr^{2} \!+\! 18\lambda M^{2}r\!+\! 18M^{3}} {r^{3}(\lambda r \!+\! 3M)^2}~~,
\eeq
being $ \lambda = 1/2(l - 1)(l + 2) $. The source $S^{l}(t,r)$ includes the derivative of the Dirac distribution, which 
appears in the process of building a single wave equation out of the seven even linearised wave equations:   

\beq
S^{l} = \frac{2( r- 2M ) 
\kappa }{r^2(\lambda +1)(\lambda r+3M)}
\times 
\left \{
\frac{r (r-2M )}{2U^0}
\delta^{\prime }[r-r_u(t)]- 
\left [ \frac{r(\lambda + 1)- 3M}{2U^0} - \frac{3MU^0(r-2M)^2}{r(\lambda r+3M )} \right ]
\delta [r-r_u(t)] 
\right \}~~,
\eeq
\end{widetext}
$U^0 = \sqrt{1 - 2M/r_{u0}}/(1 - 2M/r_u)$ being the time component of the 4-velocity and $\kappa = 4m\sqrt{(2l+1)\pi}$. The {trajectory in the unperturbed SD metric $r_u(t)$ assumes different forms according to the 
initial conditions.   
For the radial infall of a particle starting
from rest at finite distance $r_{u0}$, $r_{u}(t)$ is the inverse function of: 

\begin{widetext}
\[
\frac{t(r_u)}{2M}= \sqrt{1-\frac{2M}{r_{u0}}}\sqrt{1-\frac{r_u}{r_{u0}}}\left(\frac{r_{u0}}{2M}\right)\left(\frac{r_u}{2M}\right)^{1/2} +2{\rm{arctanh}}\left(\frac{\sqrt{\ds\frac{2M}{r_u}-\frac{2M}{r_{u0}}}}{\sqrt{\ds 1-\frac{2M}{r_{u0}}}}\right) +
\]
\beq
\sqrt{1-\frac{2M}{r_{u0}}}\left(1+\frac{4M}{r_{u0}}\right)\left(\frac{r_{u0}}{2M}\right)^{3/2}
\arctan\left(\sqrt{\frac{r_{u0}}{r}-1}\right)
\eeq
\end{widetext}
where $E=\sqrt{1-2M/r_{u0}}$. 

For an infalling mass from infinity at zero velocity, the energy radiated to infinity for all modes \cite{daruprpr71} is given by (in physical units):

\beq
\sum_l E^l_{rad} = 0.0104 \frac{m^2c^2}{M}~~,
\eeq
while most of the energy is emitted below the frequency:

\beq
f_m = 0.08 \frac {c^3}{GM}~~.
\eeq
Up to $94 \%$ of the energy is radiated between $8M$ and $2M$ and $90 \%$ of it in the quadrupole mode. Obviously 
the above expressions would sensibly vary in case of an initial relativistic velocity.   

For the SD black hole perturbations, there have been successive corrections of the basic equations, the last being pointed out by Sago, Nakano and Sasaki \cite{sanasa03}. 
Sago {\it et al.} have corrected the ZM equations (a minus sign missing in all right-hand side terms) but introduced a wrong definition of the scalar product leading to errors in the coefficients of the energy-momentum tensor \cite{na10}. 
The expressions herein correspond to those in \cite{sp11}, where some of the errors of previously published literature on radial fall are indicated. 

\section{The jump conditions}

It has been indicated by two different heuristic arguments \cite{lo00, lona09} that even metric perturbations, for radial fall in the RW gauge, should belong to the $C^0$ continuity class at the position of
the particle. One argument \cite{lo00} is based on the integration in $r$ of the Hamiltonian constraint, which is the $tt$ component of the 
Einstein equations (Eq. (C7a) in \cite{ze70c}); the other \cite{lona09} 
on the structure of selected even perturbation equations. Following the latter, Lousto and Nakano evince the continuity of the even perturbations and afterwards impose such continuity to derive the jump conditions on the wave function and its derivatives for $l = 2$, notably starting from the ZM equation.    

In this section, we instead provide an analysis {\it vis \`{a} vis} the jump conditions that the wave function and its derivatives have to satisfy for guaranteeing the continuity of the perturbations at the position of the particle. This approach \cite{aosp1011} is based on the solutions of the ZM equation, not on the equation itself. The jump conditions found herein are applicable to all modes.  Incidentally, the jump conditions were previously mentioned by Sopuerta and Laguna \cite{sola06}, Haas \cite{ha07}, and Field {\it et al.} \cite{fihela09} in a different context. 
In the forthcoming paper \cite{aosp11}, the jump conditions for generic orbits are based on the analysis of the odd and even wave equations and are determined despite the lack of continuity of the perturbations.   

In the radial case, only three independent perturbations survive, namely $K^l$, $H^l_0=H^l_2$, $H^l_1$. 
The inverse relations for those perturbations, as functions of the wave function and its derivatives, are given by (we drop henceforth the $l$ index): 

\begin{widetext}
\beq
K=\frac{6M^2+3M\lambda r+\lambda (\lambda +1)r^2}
{r^2(\lambda r+3M)}\Psi
+
\left( 1-\frac{2M}r\right) \,\Psi_{,r} -\frac{
\kappa \ U^0(r-2M)^2}{(\lambda +1)(\lambda r+3M)r}\delta~~,
\label{eq:K}
\eeq

\[ 
H_2=-\frac{9M^3+9\lambda M^2r+
3\lambda ^2Mr^2+\lambda ^2(\lambda +1)r^3}{
r^2(\lambda r+3M)^2}\,\Psi   
+
\frac{3M^2-\lambda Mr+\lambda r^2}{r(\lambda r+ 3M)}\Psi_{,r}
+ 
\]
\beq
(r-2M)\Psi_{,rr} 
+ 
\frac{\kappa U^0(r-2M)(\lambda ^2r^2+2\lambda Mr-3Mr+3M^2)}{r (\lambda +1)(\lambda r+3M)^2}\delta 
-
\frac{\kappa U^0(r-2M)^2}{
(\lambda +1)(\lambda r+3M)}\delta'~~, 
\label{eq:H02}
\eeq

\beq
H_1 = \frac{\lambda r^2-3M\lambda
r-3M^2}{
\left( r-2M\right) (\lambda r+3M)}{\Psi_{,t}} + r\Psi_{,tr} 
- 
\frac{\kappa \ U^0\stackrel{.}{z}_u(\lambda r+M)}{(\lambda +1)(\lambda
r+3M)}\delta 
+
\frac{\kappa \ U^0\stackrel{.}{z}_u r(r-2M)}{(\lambda
+1)(\lambda r+3M)}\delta'~~, 
\label{eq:H1}
\eeq
where $\delta = \delta\left[r-r_u(t)\right ]$ and $\delta' = \delta'\left[r-r_u(t)\right ]$.
\end{widetext}

From the visual inspection of the ZM wave equation (\ref{eq:rwz*}), it is evinced that the wave function $\Psi$ is of $C^{-1}$ continuity class, since the second derivative of the wave function is proportional to the first derivative 
of the Dirac distribution
(incidentally, the concept of $C^{-1}$ continuity class element, like a Heaviside step distribution, 
may be pragmatically introduced as an element of a class of functions which after integration transforms into an element belonging to the $C^{0}$ class of functions). Thus, the wave function and its derivatives can be written as: 

\beq 
\Psi=\!\Psi^+(t,r)~\Theta_1+\Psi^-(t,r)~\Theta_2~~,
\label{psi-i}
\eeq

\beq
\Psi_{,r} \!=\! \Psi^+_{,r}\Theta_1 + \Psi^-_{,r} \Theta_2 + \left(\Psi^+ \!-\!\Psi^-\right) \delta~~,
\label{psir-i}
\eeq

\beq
\Psi_{,rr} \!=\! \Psi^+_{,rr}\Theta_1 \! + \! \Psi^-_{,rr} \Theta_2 \! + \! 
\left( \Psi^+_{,r} \!-\! \Psi^-_{,r} \right) \delta \! + \! 
\left(\Psi^+ \!-\! \Psi^-\right)_{r_u}\!\! \delta'~~, 
\label{psirr-i}
 \eeq
 
\beq
\Psi_{,t} \!=\! \Psi^+_{,t}\Theta_1 \! + \! \Psi^-_{,t} \Theta_2\!-\! \left(\Psi^+ \!-\!\Psi^-\right) \dot{r}_u \delta~~,
\label{psit-i}
\eeq

\beq
\Psi_{,tr} \!=\! 
 \Psi^+_{,tr}\Theta_1 \! + \! \Psi^-_{,tr} \Theta_2 \! + \! \left(\Psi^+_{,t} \!-\!\Psi^-_{,t}\right) \delta
  \!-\! \left(\Psi^+ \!-\!\Psi^-\right)_{r_u}\!\! \dot{r}_u \delta'~~,
\label{psitr-i}
 \eeq
where $\Theta_1 = \Theta\left[r-r_u(t)\right ]$, and $\Theta_2 = \Theta\left[r_u(t) - r \right ]$ are two Heaviside step distributions. For Eqs. (\ref{psirr-i},\ref{psitr-i}), a property of the Dirac delta distribution at the position of the particle, namely $f(r)
\delta'[r-r_u(t)]=f|_{r_u(t)} \delta'[r-r_u(t)] - f'|_{r_u(t)} \delta[r-r_u(t)]$, has been used.  

The wave function $\Psi$ and its derivatives are to be replaced in Eqs. (\ref{eq:K}-\ref{eq:H1}). At the position of the particle, it is wished that the combination of all discontinuities renders the perturbations of $C^0$ class. Thus, 
the coefficients of $\Theta_1$ must be equal to the coefficients of $\Theta_2$, while 
the coefficients of $\delta$ and $\delta '$ must vanish separately. 
To this end, the perturbations are expressed in an implicit form: 

 \beq 
 K =  f_1(r) \Psi + f_2(r)\Psi_{,r} + f_3(r) \delta~~,
 \label{eq:KsymbolPsi} 
 \eeq

 \beq 
 H_2 = f_4(r) \Psi + f_5(r)\Psi_{,r} + f_6(r)\Psi_{,rr} + f_7(r)\delta+ f_8(r)\delta'~~,
 \label{eq:H02symbolPsi}
 \eeq

 \beq 
 H_1 = f_9(r)\Psi_{,t} + f_{10}(r)\Psi_{,tr} + f_{11}(r)\delta+ f_{12}(r)\delta'~~,
 \label{eq:H1symbolPsi}
 \eeq
 where the definitions of the $f$ functions may be easily drawn by visual inspection of Eqs. (\ref{eq:K}-\ref{eq:H1}). 
The jump conditions are obtained via Eq. (\ref{eq:KsymbolPsi}): 

 \beq
\left ( \Psi^+ - \Psi^- \right )_{r_u} = - \frac{f_3}{f_2}~~,
\label{eq:condkpsi} 
\eeq

\beq
\left (\Psi^+_{,r} - \Psi^-_{,r} \right )_{r_u}= \frac{f_1 f_3}{f_2^2}~~,
\label{eq:condkpsir} 
 \eeq
via Eq. (\ref{eq:H02symbolPsi}): 
 
 \beq
 \left (\Psi^+ - \Psi^- \right )_{r_u}= - \frac{f_8}{f_6}~~,
\label{eq:condh02psi} 
\eeq
 
\beq
\left ( \Psi^+_{,r} - \Psi^-_{,r} \right )_{r_u}= \frac{1}{f_6}\left( \frac{f_5 f_8}{f_6} - f_7 + f_{8,r}- \frac{f_{6,r} f_8}{f_6}\right )~~,
\label{eq:condh02psir} 
 \eeq
and finally via Eq. (\ref{eq:H1symbolPsi}):

 \beq
\left ( \Psi^+ - \Psi^- \right )_{r_u}=  \frac{f_{12}}{\dot{r}_u f_{10}}~~. 
 \label{eq:condh1psi} 
 \eeq
The jump conditions set by Eqs. (\ref{eq:condkpsi}), (\ref{eq:condh02psi}) and (\ref{eq:condh1psi}) are equivalent, as those set by Eqs. (\ref{eq:condkpsir}) and (\ref{eq:condh02psir}).
The equivalence shows the consistency of the conditions and the compliance to the continuity requirement. 
The first $t$, second $r$, second mixed $t,r$ derivative jump conditions (coming from $H_2$ and $H_1$) are given by:

 \beq
 \left ( \Psi^+_{,rr} - \Psi^-_{,rr} \right )_{r_u} = - \frac{f_4 \left(\Psi^+ - \Psi^-\right) + f_5\left(\Psi^+_{,r} - \Psi^-_{,r}\right)}{f_6}~~,
\label{psirr} 
\eeq

 \[
\left ( \Psi^+_{,t} - \Psi^-_{,t} \right )_{r_u} =  \frac{d\left(\Psi^+ - \Psi^-\right)}{dt}-\left(\Psi^+_{,r} - \Psi^-_{,r} \right)\dot{r}_{u} =
\]
\beq
\frac{\left(f_{9} - f_{10,r} \right )\dot{r}_{u} \left(\Psi^+ - \Psi^-\right) - f_{11} + f_{12,r}}{f_{10}}~~,
\label{psit} 
\eeq

 \beq
\left ( \Psi^+_{,tr} - \Psi^-_{,tr} \right )_{r_u} = - \frac{f_{9}\left(\Psi^+_{,t} - \Psi^-_{,t}\right)}{f_{10}}~~.
\label{psitr} 
\eeq

In explicit form, the jump conditions become:

\begin{widetext}

\beq
\left (\Psi^+ - \Psi^- \right )_{r_u} =
\frac{\kappa E r_u}{(\lambda +1) (3 M+\lambda   r_u)}~~,
\label{eq.jump.psi}
\eeq

\beq
\left (\Psi^+_{,r} - \Psi^-_{,r}\right )_{r_u} =
\frac{\kappa E \left[6 M^2+3 M \lambda  r_u+\lambda 
   (\lambda +1) r_u^2\right]}{(\lambda +1) (2
   M-r_u) (3 M+\lambda  r_u)^2}~~,
\eeq

\beq
\left (\Psi^+_{,rr} - \Psi^-_{,rr}\right )_{r_u} =
-\frac{\kappa E \left[3 M^3 (5 \lambda -3)+6 M^2 \lambda (\lambda -3)
    r_u+3 M \lambda ^2
(\lambda -1)    r_u^2-2 \lambda ^2 (\lambda +1)
   r_u^3\right]}{(\lambda +1) (2M-r_u)^2 (3
   M+\lambda  r_u)^3}~~,
\eeq

\beq
\left (\Psi^+_{,t} - \Psi^-_{,t}\right )_{r_u} = 
-\frac{\kappa E r_u \dot{r}_{u}}{(2 M - r_u) (3
   M+\lambda r_u)}~~,
\label{psitexp} 
\eeq

\beq
\left ( \Psi^+_{,tr} - \Psi^-_{,tr}\right )_{r_u} = 
\frac{\kappa E \left(3 M^2+3 M \lambda 
   r_u - \lambda  r_u^2\right)\dot{r}_{u}}{(2M - r_u)^2 (3 M+\lambda  r_u)^2}~~.
\label{psitrexp} 
\eeq
\end{widetext}

Having established the jump conditions for any value of $l$ at first order, we now turn to the presentation of the integration scheme. It is remarked that first derivatives of the jump conditions suffice to a scheme at first order.

\section{The numerical method}

Our integration domain is discretised using a two-dimensional uniform mesh based on the coordinates $t$, $r^*$ as depicted in Fig. \ref{fig.mesh}, where $-450 <r^*/2M<1000$  and $0<t<t_{end}$, with $t_{end}$ greater than the falling time. Each cell has an area of $2 h^2$ where $\sqrt{2} h$ is the dimension of an edge of the cell.
The evolution scheme starts with the initial data approach by Martel and Poisson \cite{mapo02}.

\begin{figure}[t]
\begin{center}
\includegraphics[width=2.2in]{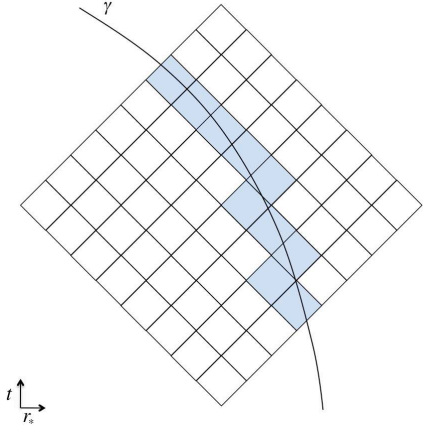}
\end{center}
\caption{Numerical domain: a staggered 1+1-dimensional mesh in $t$,$r^*$ coordinates. The geodesic trajectory of the particle is represented by the continued line $\gamma$. The shadowed cells are those crossed by the particle. 
}
\label{fig.mesh}
\end{figure}

\subsection{The numerical scheme}

The jump conditions determined in the previous section are functions of the $t,r$ variables, but for convenience in the following computations, they are transformed into functions of $t,r^*$. 
The integration method considers cells belonging to two groups: 
for cells which are never crossed by the particle, the integrating approach is drawn by the LPMP method, whereas for cells which are crossed, we propose our strategy.  

There are four sub-cases representing the different particle trajectories inside the cell, see Figs. (\ref{fig.cell.princ} - \ref{fig.case4}). We define $\alpha, \beta, \gamma, \delta$ the four vertices of the diamond centred on $\sigma$ and initially consider the world line crossing the cell as displayed in Fig. (\ref{fig.cell.princ}). The trajectory crosses the line $\beta$-$\delta$ at the point $a$ and the line $\alpha$-$\gamma$ at the point $b$. We also define the shift $\epsilon_a$ and the lapse $\epsilon_b$ as $\epsilon_a=\text{min} \left\{ r^*_a-r^*_\beta, r^*_\delta-r^*_a \right\}$, $\epsilon_b=\text{min} \left\{ t_\alpha-t_b , t_b-t_\gamma \right\}$, respectively. The jump conditions at the $a, b$ coordinates express six analytical equations:

\begin{figure}[t]
\begin{center}
\includegraphics[width=1.8in]{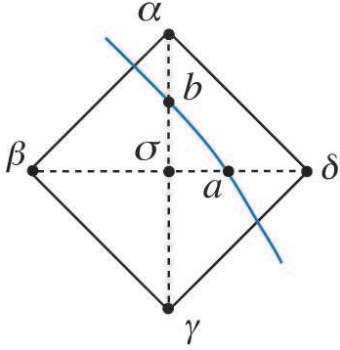}
\end{center}
\caption{Cell crossed by the particle. The particle crosses the segment $\left[ \sigma \delta \right]$ at the point $a$ and the segment $\left[ \sigma \alpha \right]$ at the point $b$.}
\label{fig.cell.princ}
\end{figure}

\beq
\left(\Psi^+ - \Psi^-\right )_a = [\Psi]_a~~,
\label{eq.jump.at.a}
\eeq

\beq
\left(\Psi^+ - \Psi^-\right )_b = [\Psi]_b~~,
\label{eq.jump.at.b}
\eeq

\beq
\left( \Psi_{,r^*}^+ - \Psi_{,r^*}^-\right )_a = [\Psi_{,r^*}]_a~~, 
\label{eq.jump.rstar.at.a}
\eeq

\beq
\left( \Psi_{,r^*}^+ - \Psi_{,r^*}^-\right )_b = [\Psi_{,r^*}]_b~~, 
\label{eq.jump.rstar.at.b}
\eeq

\beq 
\left( \Psi_{,t}^+ - \Psi_{,t}^-\right )_a = [\Psi_{,t}]_a~~, 
\label{eq.jump.t.at.a}
\eeq

\beq 
\left( \Psi_{,t}^+ - \Psi_{,t}^-\right )_b = [\Psi_{,t}]_b~~. 
\label{eq.jump.t.at.b}
\eeq

We maintain the superscript notation for which the wave function values between the world line and infinity are noted by the 
plus sign, and conversely by the minus sign for the values between the world line and the horizon (incidentally, such  notation would also be applicable to the LPMP method).    

Using a first order expansion, we get a set of six numerical equations. For case 1, see Fig. (\ref{fig.cell.princ}), they are: 

\beq
\Psi^+_\alpha   = \Psi^+   (t_b + \epsilon_b,  r^*_b) = 
\Psi^+_b + \epsilon_b \left.\Psi^+_{,t}\right|_b~~, 
\label{psi+alpha}
\eeq

\beq
\Psi^-_\sigma    = \Psi^-   (t_b - (h - \epsilon_b), r^*_b) = 
\Psi^-_b - (h -  \epsilon_b)\left.\Psi^-_{, t}\right|_b~~, 
\label{psi-sigma-uno}
\eeq

\beq
\Psi^-_\gamma    = \Psi^-(t_b - 2h + \epsilon_b, r^*_b) = \Psi^-   (t_\sigma - h, r^*_b) =
\Psi^-_\sigma - h \left.\Psi^-_{, t}\right|_\sigma~~, 
\label{psi-gamma}
\eeq

\beq
\Psi^+_\delta   = \Psi^+   (t_a, r^*_a + \epsilon_a) = 
\Psi_a^+ + \epsilon_a \left. \Psi^+_{, r^*}\right|_a~~,
\label{psi+delta}
\eeq

\beq
\Psi^-_\sigma   = \Psi^-   (t_a, r^*_a - (h - \epsilon_a)) = \Psi^-_a - (h-\epsilon_a) \left. \Psi^-_{, r^*}\right|_a~~,
\label{psi-sigma-due}
\eeq

\beq
\Psi^-_\beta    = \Psi^-(t_a, r^*_a - 2 h +  \epsilon_a) = \Psi^-   (t_\sigma, r^*_\sigma - h) = 
\Psi^-_\sigma - h \left. \Psi^-_{, r^*}\right|_\sigma~~.
\label{psi-beta}
\eeq

Our aim is the determination of the value of $\Psi^+_\alpha$, knowing those of $\Psi^-_\beta$, $\Psi^-_\gamma$, $\Psi^+_\delta$, $\epsilon_a$, $\epsilon_b$, $\left[\Psi \right]_{a,b}$, $\left[\Psi_{,r} \right]_{a,b}$ and $\left[\Psi_{,t} \right]_{a,b}$
To this end, we subtract Eq. (\ref{psi-sigma-uno}) from Eq. (\ref{psi+alpha}), Eq. (\ref{psi-sigma-due}) from Eq. (\ref{psi+delta}) and obtain, respectively:

\beq
\Psi^+_\alpha = \Psi^-_\sigma + [\Psi]_b +  \epsilon_b [\Psi_{,t}]_b + h \left.\Psi^-_{,t}\right|_b~~,
\label{psi+alpha-nuovo}
\eeq

\beq
\Psi^+_\delta = \Psi^-_\sigma + [\Psi]_a +  \epsilon_a [\Psi_{,r^*}]_a + h \left.\Psi^-_{,r^*}\right|_a~~.
\label{psi+delta-nuovo}
\eeq

Summing Eq. (\ref{psi-gamma}) and Eq. (\ref{psi+alpha-nuovo}), Eq. (\ref{psi-beta}) and Eq. (\ref{psi+delta-nuovo}), and combining the results, we finally get:

\beq
\Psi^+_\alpha = \Psi^-_\beta - \Psi^-_\gamma + \Psi^+_\delta - \left[\Psi \right]_a + \left[\Psi \right]_b 
- \epsilon_a \left[\Psi_{,r^*} \right]_a + \epsilon_b \left[\Psi_{,t} \right]_b~~. 
\label{psi+alpha-case1}
\eeq

We thus have obtained, without direct integration of the singular source, the value of the upper node. Furthermore, the latter depends solely  on analytic expressions, and obviously on the other values at the cell corners. Similar relations are found for the other three cases. For case 2 (Fig. \ref{fig.case2}), we obtain:

\beq
\Psi^+_\alpha = \Psi^-_\beta - \Psi^-_\gamma + \Psi^+_\delta + \left[\Psi \right]_a - \left[\Psi \right]_b 
- \epsilon_a \left[\Psi_{,r^*} \right]_a + \epsilon_b \left[\Psi_{,t} \right]_b~~;
\label{psi+alpha-case2}
\eeq
for case 3 (Fig. \ref{fig.case3}):

\beq
\Psi^-_\alpha = \Psi^-_\beta - \Psi^-_\gamma + \Psi^+_\delta - \left[\Psi \right]_a 
- \epsilon_a \left[\Psi_{,r^*} \right]_a~~;
\label{psi-alpha-case3}
\eeq

for case 4 (Fig. \ref{fig.case4}):
\beq
\Psi^+_\alpha = \Psi^-_\beta - \Psi^+_\gamma + \Psi^+_\delta + \left[\Psi \right]_a 
- \epsilon_a \left[\Psi_{,r^*} \right]_a~~.
\label{psi+alpha-case4}
\eeq

In a concise form, the four cases might be represented by a single expression (where $\Psi_\alpha$ stands for $\Psi^+_\alpha$ or $\Psi^-_\alpha$, and $\Psi_\gamma$ stands for $\Psi^+_\gamma$ or $\Psi^-_\gamma$):   

\begin{widetext}
\beq
\Psi_\alpha = \Psi^-_\beta - \Psi_\gamma + \Psi^+_\delta \mp \left\{\left[\Psi \right]_a - B \left[\Psi \right]_b\right\} 
- \epsilon_a \left[\Psi_{,r^*} \right]_a + B \epsilon_b \left[\Psi_{,t} \right]_b~~,
\label{psi+alpha-case1,2,3,4}
\eeq
where the upper sign holds when $r^*_a > r^*_\sigma $ and the lower when $r^*_a < r^*_\sigma $, $B = 0$ if the particle does not cross the $\alpha$-$\gamma$ line, and $B = 1$ otherwise. 

\end{widetext}

\begin{figure}[t]
\begin{center}
\includegraphics[width=1.8in]{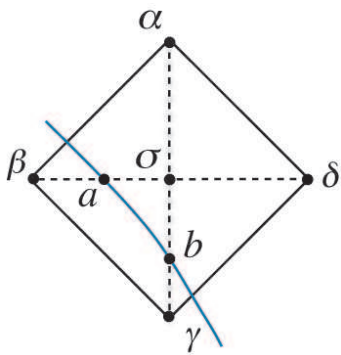}
\end{center}
\caption{Cell crossed by the particle. The particle crosses the segment $\left[ \sigma \gamma \right]$ at the point $b$ and the segment $\left[ \sigma \beta \right]$ at the point $a$.}
\label{fig.case2}
\end{figure}

\begin{figure}[t]
\begin{center}
\includegraphics[width=1.8in]{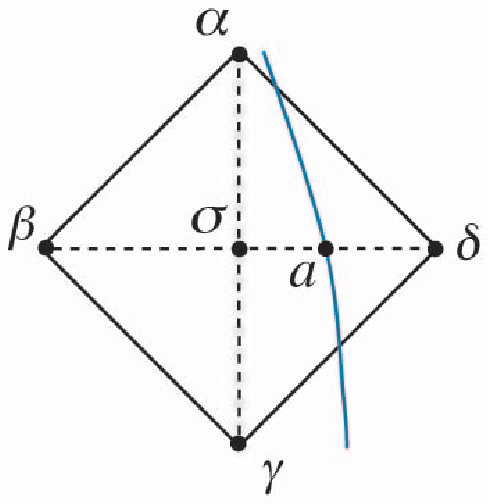}
\end{center}
\caption{Cell crossed by the particle. The particle crosses the segment $\left[ \sigma \delta \right]$ at the point $a$.}
\label{fig.case3}
\end{figure}

\begin{figure}[t]
\begin{center}
\includegraphics[width=1.8in]{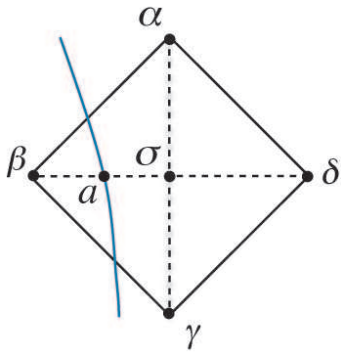}
\end{center}
\caption{Cell crossed by the particle. The particle crosses the segment $\left[ \sigma \beta \right]$ at the point $a$.}
\label{fig.case4}
\end{figure}


\section{Radiated wave forms at infinity and the wave function at the particle}

We confirm the existing wave forms previously published by Lousto and Price \cite{lopr97b}, and Martel and Poisson \cite{mapo02}. 
In this section, the results from a code based on the LPMP method (full second order) are compared to those obtained by the 
-I- method (first order for the filled cells, as presented herein, and LPMP-like for empty cells). In spite of the partially different order of the two codes, the difference between the wave forms computed by the two methods is marginal. Incidentally, this occurs in absence of a recognised standard to which refer different numerical approaches.    
 
Figs. \ref{5_superp}, \ref{20_superp} show the $l=2$ wave forms for the LPMP and -I- methods, for a particle falling radially, with zero initial velocity,  from $r_{u0}/2M=5$ and $r_{u0}/2M=20$, respectively, in units $u = t - r^*$, $(2M/m)\Psi$.} The initial data are given by the minimal ($ \alpha = 1$) initial data condition $H_2 = \alpha K$ for both cases (the parameter $\alpha$, introduced in \cite{mapo02}, measures
the amount of radiation present on the initial hypersurface and for $\alpha = 1$, the initial metric is conformally flat).

\begin{figure}[t]
\begin{center}
\includegraphics[width=3.7in]{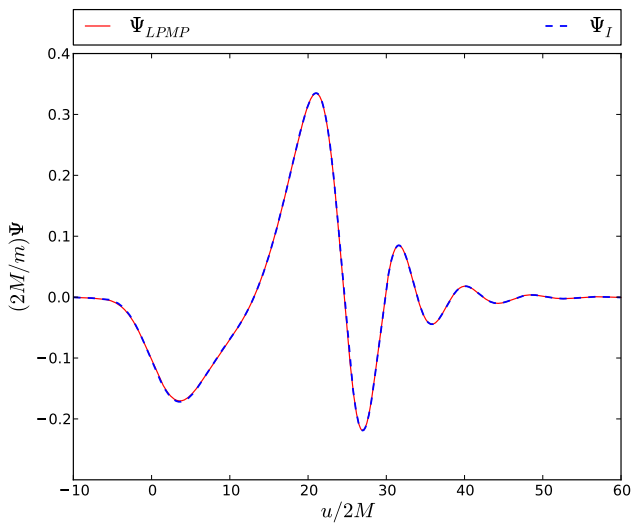}
\end{center}
\caption{Radiated $l=2$ wave form for a particle falling radially from $r_{u0}/2M=5$  with zero initial velocity, in 
units $u = t - r^*$, $(2M/m)\Psi$. The dashed and solid line represent the -I- and LPMP methods, respectively.}
\label{5_superp}
\end{figure}

\begin{figure}[t]
\begin{center}
\includegraphics[width=3.7in]{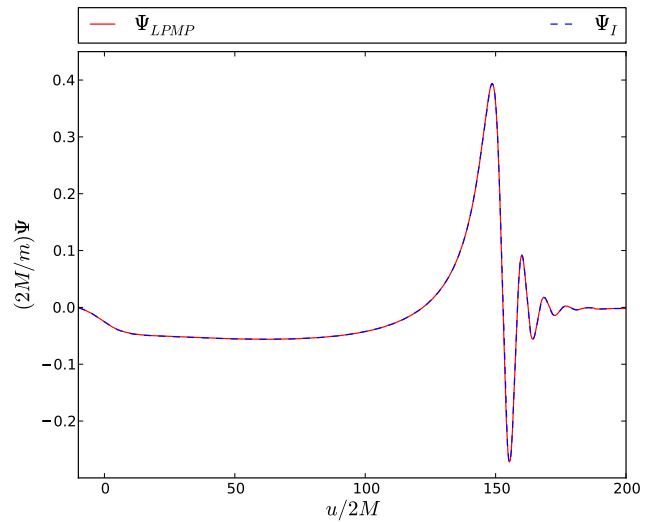}
\end{center}
\caption{ Radiated $l=2$ wave form for a particle falling radially from $r_{u0}/2M=20$ with zero initial velocity, in units $u = t - r^*$, $(2M/m)\Psi$. The dashed and solid line represent the -I- and LPMP methods, respectively.}
\label{20_superp}
\end{figure}

\begin{figure}[t]
\begin{center}
\includegraphics[width=3.7in]{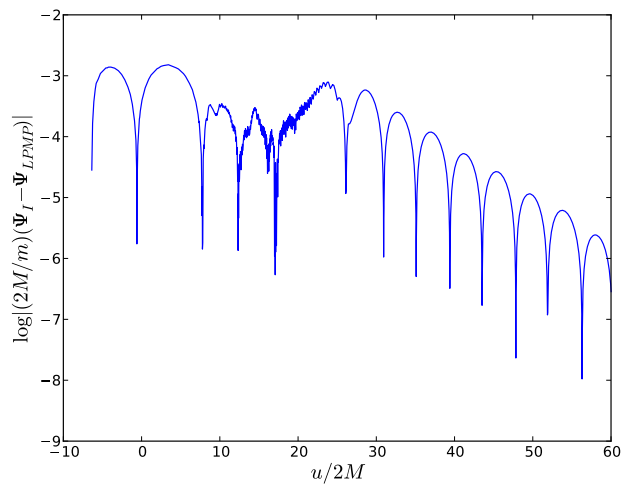}
\end{center}
\caption{Logarithmic (absolute) difference plot for a particle falling radially from $r_{u0}/2M=5$ with zero initial velocity, in units $u = t - r^*$, $(2M/m)\Psi$, between the -I- and LPMP methods ($l=2$).}
\label{5_diff_as_ns}
\end{figure}

\begin{figure}[t]
\begin{center}
\includegraphics[width=3.7in]{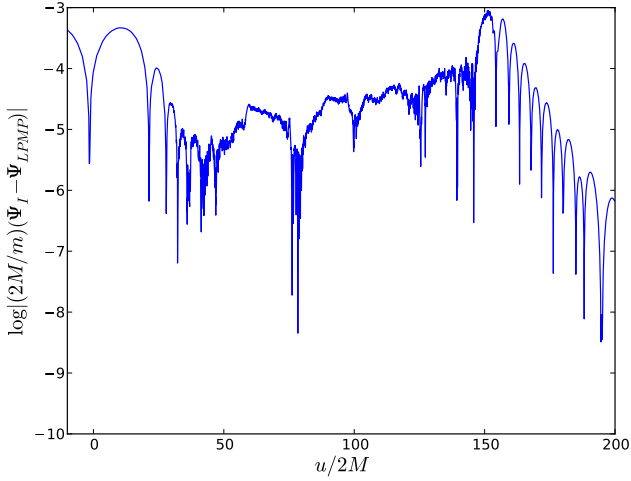}
\end{center}
\caption{Logarithmic (absolute) difference plot for a particle falling radially from $r_{u0}/2M=20$ with zero initial velocity, in units $u = t - r^*$, $(2M/m)\Psi$, between the -I- and LPMP methods ($l=2$).}
\label{20_diff_as_ns}
\end{figure}

Figs. \ref{5_diff_as_ns}, \ref{20_diff_as_ns} show the logarithmic (absolute) difference for the $l=2$ wave forms between the LPMP and -I- methods, for a particle falling radially, with zero initial velocity,  from $r_{u0}/2M=5$ and $r_{u0}/2M=20$, respectively. It is evinced that the normalised wave function amplitudes differ of an amount between $10^{-3}$ and $10^{-8.5}$, in normalised units.

\begin{figure}[t]
\begin{center}
\includegraphics[width=3.7in]{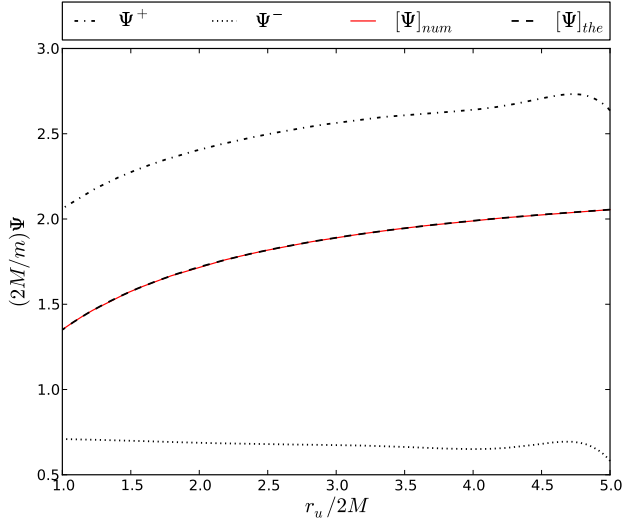}
\end{center}
\caption{Wave function ($\Psi^+$, dashed-dotted line, and $\Psi^-$, dotted line) at the particle position for a particle falling radially from $r_{u0}/2M=5$ with zero initial velocity ($l=2$), in units $u = t - r^*$, $(2M/m)\Psi$. The numerical jump condition $[\Psi]_{num}$, solid line, and the theoretical jump condition $[\Psi]_{the}$, dashed line Eq. (\ref{eq.jump.psi}), are shown.}
\label{5_jump}
\end{figure}

\begin{figure}[t]
\begin{center}
\includegraphics[width=3.7in]{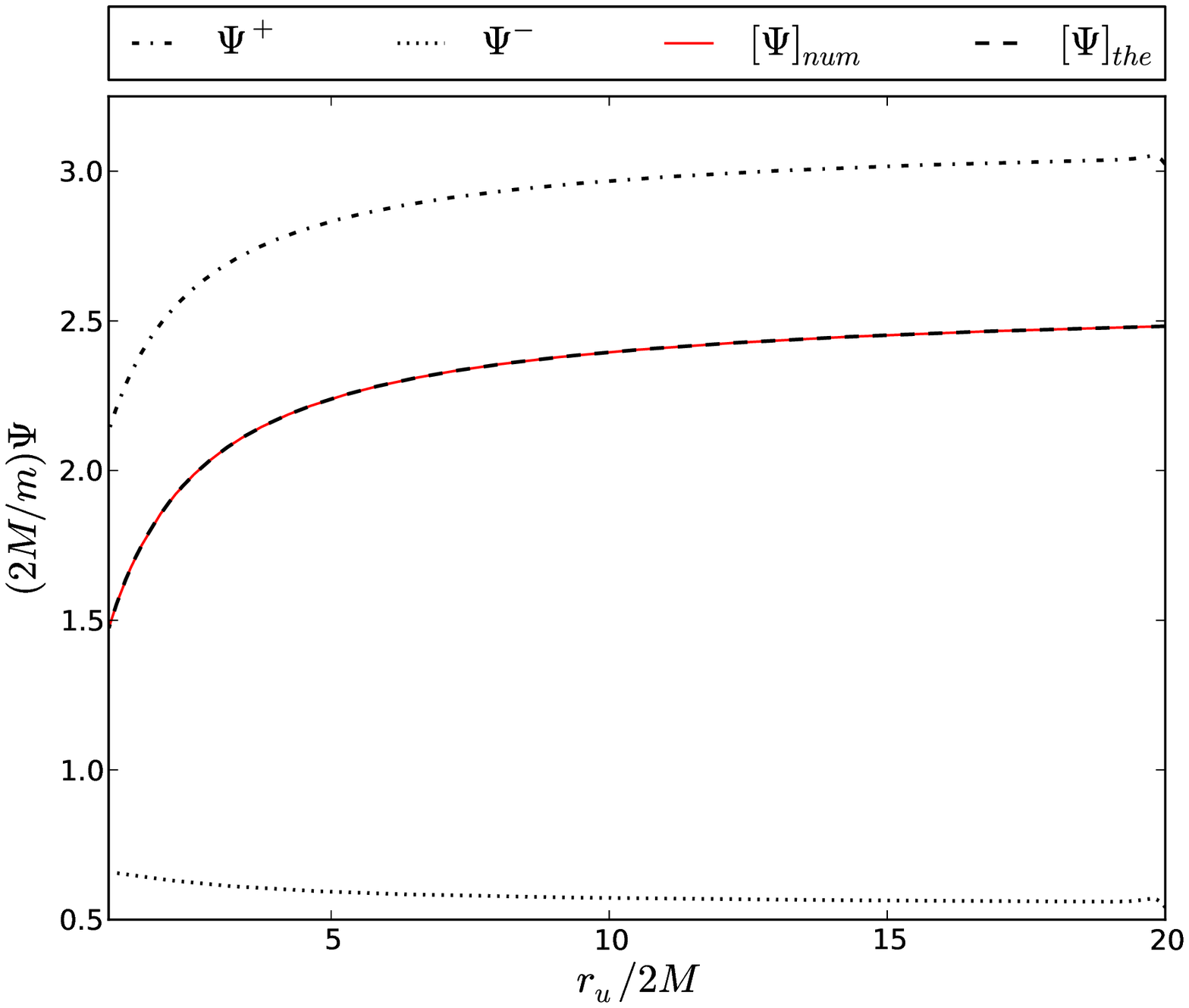}
\end{center}
\caption{Wave function ($\Psi^+$, dashed-dotted line, and $\Psi^-$, dotted line) at the particle position for a particle falling radially from $r_{u0}/2M=20$ with zero initial velocity ($l=2$), in units $u = t - r^*$, $(2M/m)\Psi$. The numerical jump condition $[\Psi]_{num}$, solid line, and the theoretical jump condition $[\Psi]_{the}$, dashed line Eq. (\ref{eq.jump.psi}), are shown.}
\label{20_jump}
\end{figure}

\begin{figure}[t]
\begin{center}
\includegraphics[width=3.7in]{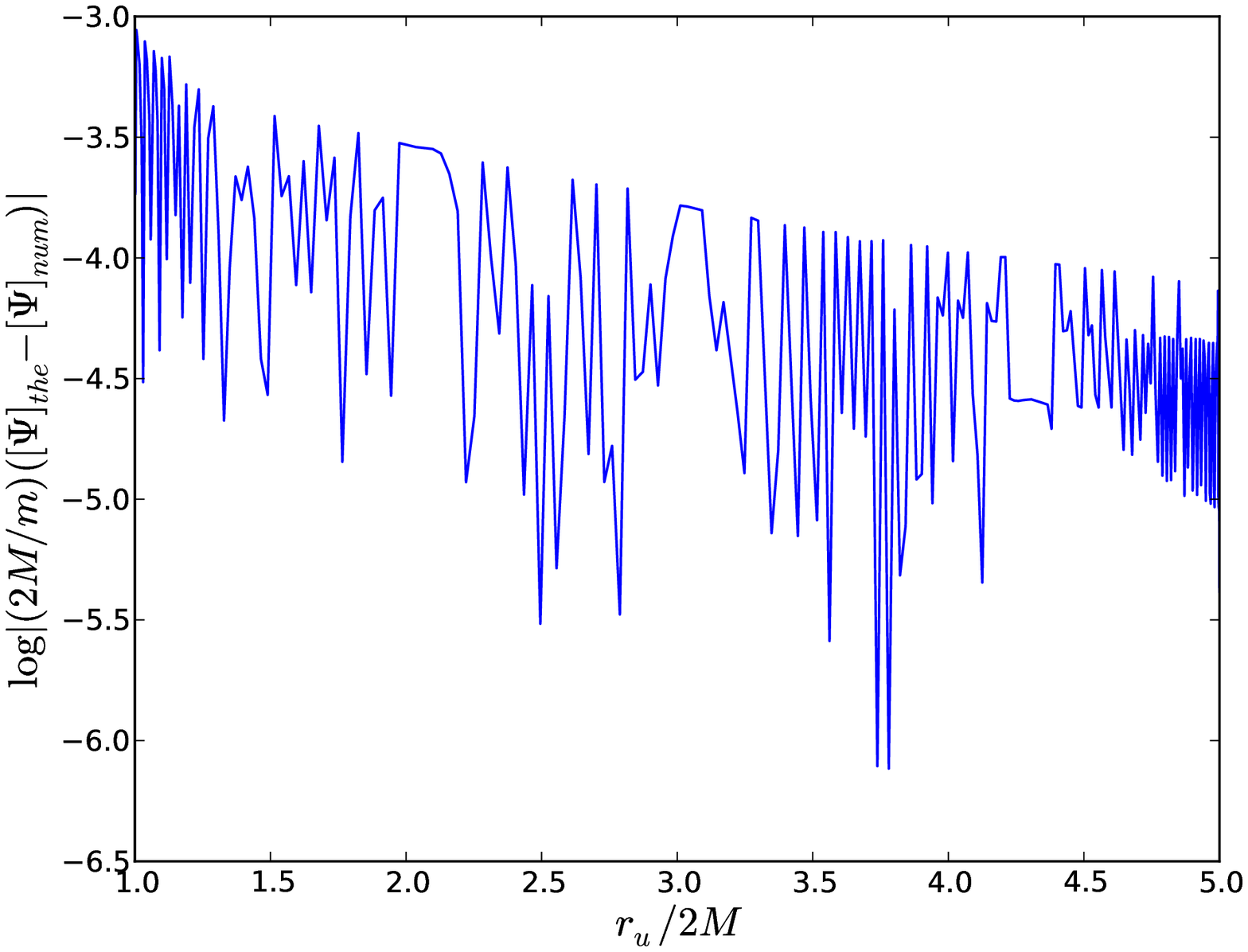}
\end{center}
\caption{Logarithmic (absolute) difference plot for a particle falling radially from $r_{u0}/2M=5$ with zero initial velocity, in units $u = t - r^*$, $(2M/m)\Psi$, at the particle position, between theoretical and numerical jump conditions ($l=2$).}
\label{5_diff_jump_noise}
\end{figure}

\begin{figure}[t]
\begin{center}
\includegraphics[width=3.7in]{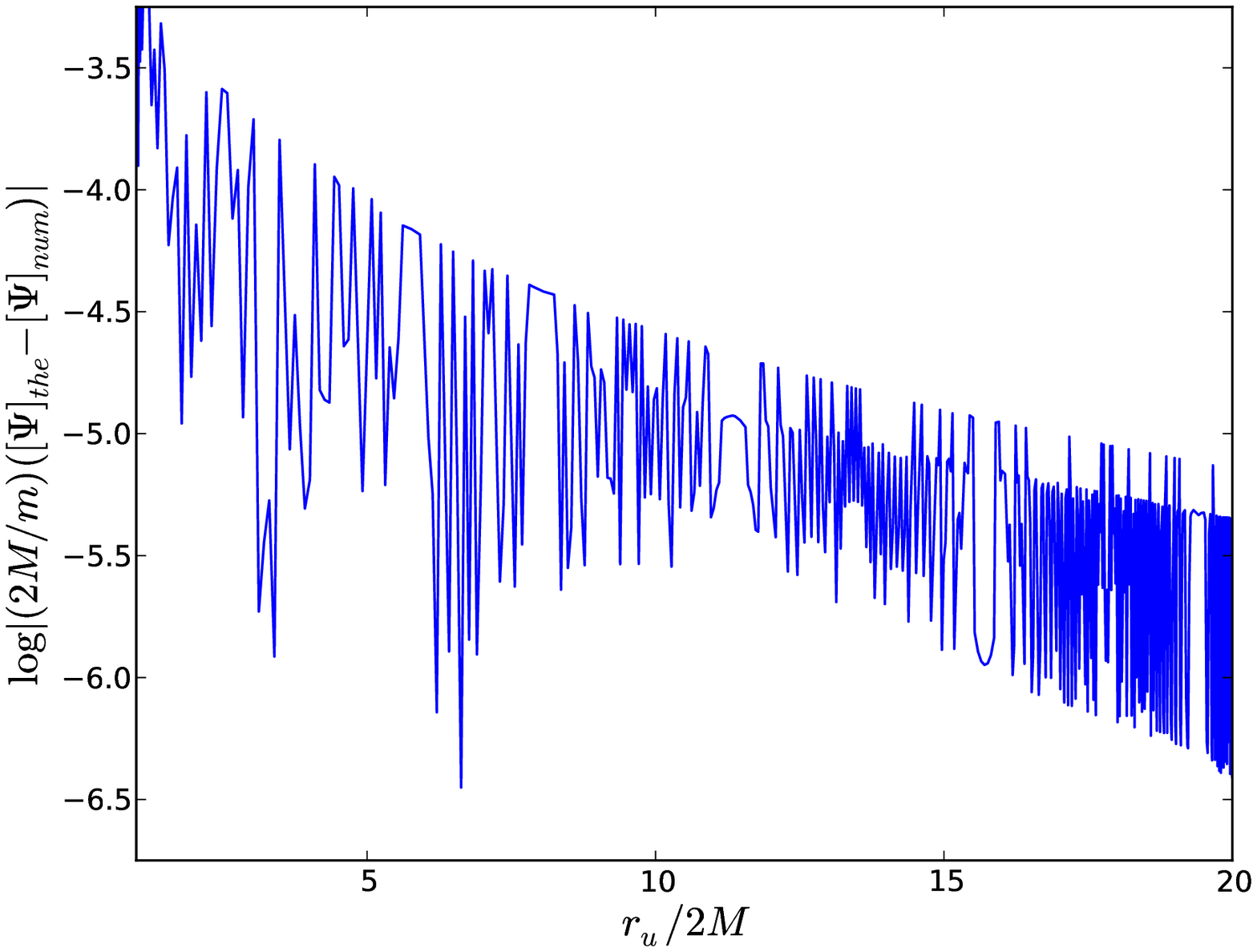}
\end{center}
\caption{Logarithmic (absolute) difference plot for a particle falling radially from $r_{u0}/2M=20$ with zero initial velocity, in units $u = t - r^*$, $(2M/m)\Psi$, at the particle position, between theoretical and numerical jump conditions ($l=2$).}
\label{20_diff_jump_noise}
\end{figure}

We show also the behaviour of the wave function at the position of the particle, Figs. (\ref{5_jump}, \ref{20_jump}). The theoretical jump condition, Eq. (\ref{eq.jump.psi}), and the numerical result well overlap at the particle position, as it may be inferred from the logarithmic (absolute) difference plots, Figs. (\ref{5_diff_jump_noise},\ref{20_diff_jump_noise}). The discrepancy in normalised units is between $10^{-3}$ and $10^{-6.5}$.

{\subsection{ Discussion}

The features of the -I- method can be summarised as follows:

\begin{itemize}
\item{ For the grid cells crossed by the particle, direct and explicit integration of the wave equation, the source term with the associated singularities and the potential, is avoided. This determines also a faster code.}
\item { Reliability is improved, since analytic expressions totally replace the numerical expressions representing the source. The terminology  ``source-free'' is due to this feature.}
\item { A higher order scheme can be written, starting from a higher order Taylor expansion 
for Eqs. (\ref{psi+alpha}-\ref{psi-beta}), and associated to this method \cite{rispaoco11}.}
\item { The applicability of the method stretches out to generic orbits. It is our concern to apply the -I- method to circular and eccentric orbits, even if these orbits are not accompanied by perturbations of $C^0$ class. The rationale poses on the following consideration. Instead of using the continuity of the perturbations to get the jump conditions, we assume that the even and odd wave equations 
are satisfied by $\Psi$, respectively $R$, being $C^{-1}$. Using this approach, we have obtained encouraging results for an eccentric orbit \cite{aosp11}.}
\item { The disadvantage of this method is represented by the four possible trajectories that a particle may follow inside a given cell, instead of the three possible paths of the LPMP method. A different labeling of the intersections between the particle world line and the cell, though, reduces the number of cases to three \cite{rispaoco11}.} 
\end{itemize}

\section{Conclusions and outlook}

We have presented a novel integration method in the time domain for the Zerili-Moncrief wave equation at first order, for cells crossed by the particle world line. 
The forward time value in the upper node of the $(t, r^*$) grid cell is obtained by the linear combination of the three preceding node values and of the analytic jump conditions. Therefore, the numerical integration does not deal any longer  with the source term, the associated singularities and the potential. The direct integration of the wave equation is circumvented. 

The wave forms at infinity confirm published results, and the wave function at the particle position shows 
a well-behaved pattern. Our final aim is the evaluation and the application of the back-action at the position of the 
particle, in a self-consistent manner. 

The indirect or source-free method has been developed to fourth order \cite{rispaoco11} and applied to generic orbits 
\cite{aosp11}.

Beyond the scenario of EMRIs, other lines of investigation may emerge. One concerns the formal relation between the algorithms of the indirect and LPMP approaches \cite{lopr97b, mapo02}. Another considers applicability of the indirect method to all wave equations with a singular source term, and which the wave function of is constrained by a set of properties.\\ 

\section*{Acknowledgements}

Discussions with St\'ephane Cordier (MAPMO-Orl\'eans), Marc-Thierry Jaekel (LPT ENS - Paris), Jos\'e Martin-Garcia (now at Wolfram Research) and Patxi Ritter (doctorate student - Orl\'eans) have clarified the jump conditions. Support from the latter in coding and producing the figures is acknowledged. John Alexander Tully, astronomer at the Observatoire de la Cote d'Azur in Nice, is thanked for reading the manuscript. Helpful comments from the referee were appreciated and implemented. The authors wish to acknowledge the FNAK (Fondation Nationale Alfred Kastler), the CJC (Conf\'ed\'eration des Jeunes Chercheurs) and all organisations which stand against discrimination of foreign researchers.

\end{document}